\documentclass{JHEP3}

\newcommand{\be}{\begin{equation}}
\newcommand{\ee}{\end{equation}}
\newcommand{\bea}{\begin{eqnarray}}
\newcommand{\eea}{\end{eqnarray}}
\newcommand{\bref}[1]{(\ref{#1})}

\newcommand{\Lmat}{\mathcal{L}_\mathrm{mat}}
\newcommand{\LGB}{\mathcal{L}_\mathrm{GB}}
\newcommand{\mm}{\, \mathrm{mm}}
\newcommand{\km}{\, \mathrm{km}}
\newcommand{\pa}{\partial}
\newcommand{\mpl}{M_\mathrm{Pl}}
\newcommand{\mph}{M_{\phi}}
\newcommand{\ms}{M_\star}
\newcommand{\dz}{\partial_z}

\title{Scalar-Tensor Gravity on a Gauss-Bonnet Brane World}
\author{Stephen C. Davis \\
Institut de Physique Th\'eorique, \'Ecole Polytechnique
F\'ed\'erale de Lausanne, \\ CH--1015 Lausanne, Switzerland \\
E-mail: \email{stephen.davis@epfl.ch}}

\abstract{
The effective four-dimensional, linearised gravity for a brane world
model with higher order curvature terms and a bulk scalar field is
analysed. Large and small distance gravitational laws are derived. The
model has a single brane embedded in a five-dimensional bulk
spacetime, and the scalar field represents the dilaton or a moduli
field. The quadratic, Gauss-Bonnet curvature term (and corresponding
higher kinetic terms for the scalar) is also included in the bulk
action. It is particularly natural to include such terms in a brane
world model. Boundary terms and junction conditions for the higher
order terms are given.

The extra terms allow additional solutions of the field equations,
which give better agreement with observational
constraints. Brans-Dicke gravity is obtained on the brane. The scalar
and tensor perturbations are affected differently by the higher
gravity terms, and this provides a way for the scalar modes to be
suppressed relative to the tensor ones. Another new (but less useful)
feature is the appearance of instabilities for some parameter ranges.
}

\keywords{eld, ctg}

\preprint{}

\begin{document}

\section{Introduction}

There has been considerable interest in the idea~\cite{earlyBW} that
we may live on  3+1 dimensional brane embedded in a 4+1 dimensional
bulk spacetime. Despite the extra dimension, it is still possible for
the effective gravitational theory on the brane to closely resemble
that which is observed in our universe. The models we will be
considering in this paper resemble the well known Randall-Sundrum II
brane world model~\cite{RSII}. In this scenario the graviton spectrum
has a zero mode which, due to the warping of the bulk space, is
localised on the brane. This is interpreted as the four-dimensional
graviton. At large distance scales the effects of the other modes are
suppressed, and something close to four-dimensional general relativity
is obtained for brane-based observers~\cite{Garriga}. Gravity will have
unconventional behaviour at short distances, and so to be compatible
with gravity experiments, `short' needs to of order $0.1\mm$ or less.

The brane world scenario is loosely motivated by string
theory. However, if such a model were actually derived from string
theory, the bulk space would contain not only gravity, but many other
fields as well. The simplest example of this would be a scalar field
such as the dilaton, or a moduli field coming from the
compactification of other extra dimensions. One of the purposes of
this paper is to investigate the implications of bulk scalar fields
for brane world gravity. For simplicity I will include just one such
field, and consider only a special class of solutions for which the
field equations can be easily solved.

Scalar-tensor gravity has previously been studied in a brane world
context~\cite{scagrav}, although most previous work has considered
only the graviton modes. However, given the strong constraints placed
on scalar modes by solar system measurements~\cite{Damour}, we should
not ignore them if we are studying a brane world which is supposed
to represent our universe.

Another natural extension of the usual brane world model is to add
higher order curvature terms to the action. One such term is the
quadratic Gauss-Bonnet (or Lanczos~\cite{Lanczos}) term
\be
\LGB = R^2 - 4 R_{ab}R^{ab} + R^{abcd} R_{abcd} \ .
\label{LGB}
\ee
Like the Einstein-Hilbert action this gives rise to field equations
which are divergence free, and do not contain any third or higher
order derivatives of the metric. They therefore have all the
desirable properties of the Einstein equation (such as a
ghost-free vacuum and conservation of energy
momentum)~\cite{Lovelock}. Note that this is the only quadratic
curvature term which has all these properties. In four dimensions its
contribution to the field equations is trivial, and so it is usually ignored. 

The implications of this term for brane world gravity have been
discussed in the
literature~\cite{Kim,RSGB,Meissner,nathalie,us2}. Note that the
derivation of the brane junction conditions is more complicated than
in the conventional brane world~\cite{RSGB,BCGB}, and this led to some
subtle mistakes in early works~\cite{Kim}.

Like the branes themselves, higher order curvature terms (and
corresponding higher order scalar kinetic terms) are also motivated by
string theory. It is therefore very natural to consider them in brane
world models. For simplicity, we will only consider the Gauss-Bonnet
term and the corresponding scalar field terms in this paper (see
ref.~\cite{Meissner} for a discussion of even higher order curvature
terms in a brane world context).

The layout of the paper is as follows: In section~\ref{sec:sol} the
details of the  model are discussed. Solutions of the field equations
are described, with and without the higher order terms in the
action. Not surprisingly, the addition of the higher order terms
produces extra solutions to the field equations. For want of a better
name we will call these the `Gauss-Bonnet branches', and the other
solutions the `Einstein branch'. Linearised brane world gravity is
reviewed in section~\ref{sec:lin}. Linear perturbations of the
`Einstein' solutions are considered in section~\ref{sec:Ein}, and the
more interesting and better behaved `Gauss-Bonnet' solutions are
analysed in section~\ref{sec:GB}. Several of the solutions are shown
to be unstable, although some of these instabilities can be removed by
adding appropriate terms to the brane part of the action. These
stabilised solutions are discussed in section~\ref{sec:Ind}. The last
section contains a summary of the results. The bulk field equations
and junction conditions (together with corresponding boundary terms)
are listed in the appendix.

\section{Bulk Solutions}
\label{sec:sol}

We will consider a five-dimensional toy model with a single,
$Z_2$-symmetric brane and one bulk scalar field, $\Phi$, which is
conformally coupled to gravity. We will be working in the
Jordan frame (in a string theory context, and when $\Phi$ is the
dilaton, this is known as the string frame). We start by considering
the first order action 
\be
\label{act1}
S_1=\frac{M^3}{2}\int d^5x \sqrt{-g} \, e^{-2\Phi}
\left\{R -4\omega (\nabla \Phi)^2 - 2\Lambda \right\} 
- M^3\int d^4x \sqrt{-h}  \, e^{-2\Phi} \left\{
2K + T \right\} \ , \hspace{0.3in} {}
\ee
where $h_{ab} = g_{ab} - n_a n_b$ is the induced metric on the brane and 
$K_{ab} = h^c{}_a \nabla_{\! c} n_b$ is the extrinsic curvature. $M$
is the five-dimensional Planck mass, $\Lambda$ is the bulk
cosmological constant, and $T$ is the brane tension. The
brane is being treated as a boundary of the two halves of the bulk
spacetime, and the Gibbons-Hawking boundary term has been included to
give a consistent action~\cite{GibHawk}. In this paper we take the
brane normal to be pointing away from the brane on both sides of the
brane. This is an unconventional choice, but it is easier to visualise.

The general second order action is
\bea
\label{act2}
&&S_2=\frac{M}{2}\int d^5x \sqrt{-g} \, e^{-2\Phi}
\left\{c_1 \LGB - 16 c_2 G_{ab} \nabla^a\Phi\nabla^b\Phi
+ 16 c_3 (\nabla \Phi)^2 \nabla^2\Phi - 16 c_4 (\nabla \Phi)^4 \right\} 
\nonumber \\  &&\hspace*{0.3 in} 
{}-M\int d^4x \sqrt{-h}  \, e^{-2\Phi} \biggl\{
\frac{4}{3} c_1 \left(3 K K_{ab}K^{ab} - 2K^a{}_b K^{bc}K_{ac} 
- K^3 -6 \widehat G^{ab}K_{ab} \right) 
\nonumber \\  &&\hspace*{1.1 in} {}
- 16 c_2 (K_{ab}- K h_{ab})D^a\Phi D^b \Phi
+ \frac{16}{3} c_3\left((\partial_n\Phi)^3 
+ 3\partial_n\Phi (D\Phi)^2\right) \biggr\} \ ,
\eea
where $\partial_n \equiv n^a \partial_a$. The caret denotes
tensors with correspond to the induced metric $h_{ab}$, and $D_a$ is
the corresponding covariant derivative.

When the scalar field is included there are a total of four possible
second order terms which produce suitable field equations (no higher
than second derivatives, etc.). For the action to be consistent, we
must include the Gauss-Bonnet equivalent of the Gibbons-Hawking
boundary term~\cite{BCGB,Myers}, as well as the corresponding
$\Phi$ dependent terms (see appendix). 

We see that the boundary parts of the above actions include terms
which are first and third order in derivatives. It is therefore
natural to also include an induced gravity action 
\be
S_\mathrm{Ind} = \frac{M^2}{2}\int d^4x \sqrt{-h}  \, e^{-2\Phi}
\left\{b_1 \widehat R + b_2 (D \Phi)^2 \right\} \ ,
\label{actind}
\ee
which is second order in derivatives. Although we will consider this
term, we will be mainly interested the implications of the other parts
of the action, \bref{act1} and \bref{act2}.

Scalar-tensor theories of gravity are also studied in the Einstein
frame, which is related to the Jordan frame by the conformal
transformation. The transformation is chosen so that $\Phi$-dependent
factor in front of Einstein-Hilbert part of the action is
cancelled. For a brane world this is either 
$g_{ab}\rightarrow e^{4\Phi/3} g_{ab}$ or $g_{ab}\rightarrow e^{2\Phi}
g_{ab}$, depending on whether we are considering the bulk~\bref{act1}
or brane~\bref{actind} part of the action. So we see that there are
actually two Einstein frames for a brane world.

The values of the coefficients $\omega$, $c_i$ and $b_i$ will be
determined by the fundamental theory from which the
actions~(\ref{act1}--\ref{actind}) are derived. One of the simplest
possibilities is to suppose that $\Phi$ is a moduli field arising from
a toroidal compactification of a higher dimensional
space. Consider a $(5+N)$-dimensional theory, whose bulk action
contains only the Einstein-Hilbert and Gauss-Bonnet terms. Take the
metric to have the form
\be
ds_{5+N}^2 = g_{ab}(x) dx^a dx^b + e^{-4\Phi(x)/N} \eta_{AB} dX^A dX^B
\ee
and then compactify the last $N$ dimensions (we have made the
simplifying assumption that these extra dimensions only have one
degree of freedom)~\cite{us2}. Comparison of field equations reveals
that such a theory will be equivalent to that obtained from the above
actions if $\omega = -1 + N^{-1}$, 
\be
c_2 = -\omega c_1 \ , \ \
c_3 = (2\omega+1)\omega c_1 \ , \ \  c_4 = -(2\omega+1)\omega^2 c_1
\label{KKcoeff}
\ee
and $b_2 = -4\omega b_1$. In the $N \to \infty$ limit, $\Phi$ can be
interpreted as the dilaton for a theory with a high degree of
symmetry~\cite{us2}. For the rest of this paper we will assume the
above relation between $c_i$, $b_i$ and $\omega$, and take
$0 \geq \omega \geq -1$. For later convenience we define $\alpha=c_1/ M^2$
and $\beta = b_1/M$. Both of these parameters will be taken to be positive.

Combining all the above contributions to the action
(\ref{act1}--\ref{actind}), and including some brane matter gives 
the total action
\be
S=S_1+S_2+S_\mathrm{Ind} - \int d^4x \sqrt{-h}  \, \Lmat \ .
\ee
The brane matter Lagrangian $\Lmat$ will be treated as a small
perturbation to the background solution. For now we will set it to zero.
 
In this paper we will be interested in perturbations of non-singular
brane world solutions of the form
\be
ds^2 = e^{-2k |z|} dx^\mu dx^\nu \eta_{\mu\nu} + dz^2 \ ,
\ee
with $\Phi = \phi_0-\sigma |z|$. The single brane is located at $z=0$,
and the bulk is taken to be $Z_2$-symmetric. Solutions of this form
are the simplest extension of the usual Randall-Sundrum
model~\cite{RSII}. Previous work~\cite{Mavromatos,Jakobek,us} on such
solutions in Gauss-Bonnet brane worlds has used different coefficients
($c_i$) for the terms in bulk action~\bref{act2}, although this did
not lead to qualitatively different results. Other, more general
classes of solutions exist~\cite{us2,Jakobek}, but we will not
consider them here.

There are problems with the above solution if $\sigma < 0$. Although
(in the Jordan frame) there are no curvature singularities, the
coupling of matter to gravity (which is proportional to $e^\Phi$) will
be divergent as $z \to \infty$. If $k > 0$ the time taken (as observed
on the brane) for a massless particle to travel from this singularity
to the brane will be infinite. However it is still possible for
signals from points arbitrarily close to the singularity (and with
arbitrarily high coupling) to reach the brane in finite time. The low
energy action we are using will not be valid near the singularity, and
so the physics of our brane universe would be dominated by effects
that we have not included in our model. In the Einstein frames (bulk
and brane), the curvature at this point becomes singular, and
furthermore the singularity will then be at a finite proper distance
from the brane.  Therefore, in order to have a consistent single brane
model, we will only consider solutions with $\sigma \geq 0$. $e^\Phi$
is then bounded above by $e^{\phi_0}$, its value on the brane.

Considering only the lowest order action~\bref{act1}, the bulk
field equations (see appendix) are solved by
\be
k = -2(\omega+1)\sigma \ .
\ee
The value of $\sigma$ will be related to the bulk cosmological
constant. We see that (for the $\sigma >0$, $\omega \geq -1$ solutions
we are considering) the sign of warp factor is the opposite of that in
the usual brane world model. As we will see in section~\ref{sec:Ein},
the negative warp factor does not give a suitable effective theory of
gravity on the brane.

When the higher order terms~\bref{act2} are included, the above
solution is still valid, and (if $\omega \neq 0$) two new branches of
solutions appear
\be
k = -\omega \sigma \pm
 \sqrt{\frac{1}{12\alpha} - \frac{\omega(\omega+2)}{3}\sigma^2}
\ee
(if $\omega=0$ then $k^2= 1/[12\alpha]$, the field equations are
degenerate and $\Phi$ is undetermined). It is possible for these
solutions to have positive warp factor, suggesting that a viable
effective gravitational theory could be obtained on the brane. These
new solutions will be referred to as the `Gauss-Bonnet branches', and
the other solution as the `Einstein branch'. It is perhaps worth
noting that for a more general choice of the coefficients in the
action~\bref{act2}, the `Einstein' solution will not still be valid
when the higher order gravity terms are switched on (see
e.g.~\cite{Mavromatos,us}).

\section{Linearised Brane Gravity}
\label{sec:lin}

We will start by reviewing linearised gravity for a brane world
without scalar fields.  By taking $\Lmat$ to be small, a perturbative
analysis can be used to determine the approximate effective
gravitational law on the brane. We will use a gauge in which the
brane is kept straight, even when matter is
present~\cite{lingrav1,lingrav2}. 

The general perturbed metric can be written as
\be
ds^2 = e^{-2k|z|}(\eta_{\mu \nu} + \gamma_{\mu \nu}) dx^\mu dx^\nu 
+ 2 v_\mu dx^\mu dz + (1+\psi) dz^2 \ .
\label{pertmet}
\ee
The quantities $\gamma_{\mu \nu}$, $v_\mu$ and $\psi$ are all
small. The perturbed brane normal is then 
$n_a = \delta^z_a \, \mathrm{sign}(z) \, (1+\psi/2)$. We use the
metric convention $\eta_{\mu \nu} = \mathrm{diag}(-1,1,1,1)$.

The lapse function, $\psi$, and the shift vector, $v_\mu$, are
determined by solving the components of the field equations which are
perpendicular to the brane. If the brane does not bend, as we are
requiring, these will not depend on the brane matter. For the
Gauss-Bonnet-Randall-Sundrum model ($k>0$) they are solved (for $z>0$) by
\be
\psi= -\frac{1}{4k} \, \dz \gamma 
\label{gaugers1}
\ee
\be
v_\mu = -\frac{1}{8k} \, \partial_\mu \gamma + B_\mu
\label{gaugers2}
\ee
\be
\partial^\mu \bar \gamma_{\mu \nu} = 0
\label{gaugers3}
\ee
where $\bar \gamma_{\mu \nu} = \gamma_{\mu \nu} - (1/4) \gamma \eta_{\mu \nu}$,
$\gamma = \eta^{\mu \nu}\gamma_{\mu \nu}$, and $B_\mu$ satisfies
$\partial^\mu B_\mu= 0$ and $\Box_4 B_\mu=0$. We can use gauge freedom to
set $B_\mu=0$.

For the above choice of gauge the remaining bulk field equations reduce to
\be
(1 - 4\alpha k^2)\left(\dz^2 -4k \dz + e^{2k z}\Box_4\right) 
\bar \gamma_{\mu\nu}=0
\label{Brs}
\ee
with $\Box_4=\eta^{\mu\nu}\partial_\mu \partial_\nu$. There is no bulk
equation for $\gamma$. The remaining gauge freedom can used to choose
its behaviour there (but not its value on the brane)~\cite{lingrav1}.

Note that the above equation is free from third or higher order
derivatives of $\bar \gamma_{\mu\nu}$, and is similar to linearised
Einstein gravity in this respect. This follows from the fact that the
Gauss-Bonnet term was specifically constructed to give a Einstein-like
theory of gravity.  If  $1 - 4\alpha k^2 < 0$ the $\bar \gamma_{\mu\nu}$ 
kinetic term in the linearised effective action will be negative, and
so the theory will have bulk ghosts~\cite{Meissner,us2}. Therefore if
the theory is to be stable on the quantum level, $k$ must be less
than $1/\sqrt{4\alpha}$.

The junction conditions at $z=0$ give
\be
2(1-4\alpha k^2)\dz \bar \gamma_{\mu\nu}
+ (\beta + 8\alpha k) \Box_4 \bar \gamma_{\mu\nu}
=-\frac{2}{M^3}\left\{S_{\mu\nu} - \frac{1}{3} 
\left(\eta_{\mu\nu} - \frac{\partial_\mu \partial_\nu}{\Box_4}\right) 
S \right\}
\label{pertbc}
\ee
and
\be
(1 + \beta k + 4\alpha k^2) \Box_4 \gamma = \frac{4k}{3M^3}S
\ee
where $S_{ab} = 2\delta \Lmat/\delta h^{ab} - h_{ab} \Lmat$ is the
energy-momentum tensor for the brane matter.

Switching to Fourier space, the bulk graviton equation~\bref{Brs} is
solved, for spacelike momenta ($p^2>0$), by
\be
\bar \gamma_{\mu\nu}(p,z) \propto e^{2k|z|} 
K_2 \left(p e^{k |z|}/k\right) \ .
\label{gsolrs}
\ee
We do not use the other solution as it diverges as $z\to \infty$.
Substituting this into the junction conditions allows
$\gamma_{\mu\nu}$, and hence the effective gravitational laws, to be
determined on the brane.

Define $\mathcal{G}_{\mu\nu}$ to be the Einstein tensor
corresponding to the perturbation $\gamma_{\mu\nu}$. 
At large distances ($p \ll k$) an asymptotic expansion of the
solutions shows that, to leading order in $p$,
$\mathcal{G}_{\mu\nu} \approx \mpl^{-2} S_{\mu\nu}$ with
\be
\mpl^2 = \frac{1+k\beta+4\alpha k^2}{k} M^3 \ .
\ee
Therefore general relativity with effective four dimensional Planck
mass $\mpl$ is obtained.

If at least one of $\alpha$ and $\beta$ is non-zero, the linearised
brane gravity at short distances ($p \gg k$) will instead be described by 
\be
\mathcal{G}_{\mu\nu} 
- 2(\eta_{\mu\nu} \Box_4 - \pa_\mu \pa_\nu) \tilde \varphi
\approx  \mpl^{-2}S_{\mu\nu}
\label{BD1}
\ee
\be
-2\Box_4 \tilde \varphi \approx \mph^{-2} S
\label{BD2}
\ee
where
\be
\mpl^2=(\beta+8\alpha k)M^3 
\ee
\be
\mph^2 =\frac{3(1 + \beta k + 4\alpha k^2)}{4(1- 4\alpha k^2)} \mpl^2 \ ,
\ee
and $\varphi \propto \gamma$. This is linearised Brans-Dicke
gravity. $\mph^{-2}$ gives the coupling strength for the
scalar degree of freedom. If Brans-Dicke theory is to be compatible with
gravitational measurements of the solar system, we need
$\mpl^2/\mph^2 < 10^{-3}$~\cite{Damour}. 

If $\alpha=\beta=0$ the short distance gravity will be very
non-standard, and this leads to strong constraints on the brane world
parameters ($1/k \lesssim 0.1 \mm$). In contrast, the short distance
constraints on scalar-tensor gravity theories, and hence the above
brane world, are quite weak ($1/k \lesssim 100 \km$ if $4\alpha k^2$
is near to 1~\cite{nathalie}).

\section{Einstein Branch}
\label{sec:Ein}

A similar analysis to that in the previous section can be applied to the
brane world solutions of section~\ref{sec:sol}. We use the perturbed
metric~\bref{pertmet}, and take $\Phi = \phi_0 - \sigma |z| +
\varphi$, with $\varphi$ small. As before, $v_\mu$ and $\psi$ are
determined by requiring that the brane remains at $z=0$. The scalar
field perturbations mix in with the trace of the metric perturbations,
and this requires a slightly different choice of gauge.

The `Gauss-Bonnet' branches will be considered in the next section. We
will begin with the `Einstein' branch, which has 
$k = -2(\omega+1)\sigma$. We are interested in the parameter range 
$0 \leq \omega \leq -1$, and so in contrast to the Randall-Sundrum model,
this solution will have negative warp factor. Not surprisingly the 
resulting brane gravity is significantly different to that in
the previous section.

It is useful to split $\gamma_{\mu\nu}$ into tensor and scalar parts
\be
\gamma_{\mu \nu} = 
\bar \gamma_{\mu \nu} + \frac{1}{4} \gamma \eta_{\mu \nu} 
+\frac{4}{3}\left(\frac{1}{4}\eta_{\mu \nu}
  -\frac{\pa_\mu \pa_\nu}{\Box_4}\right)\chi
\label{gbar}
\ee
with
\be
\chi=\frac{8k\varphi - \sigma \gamma}{4(2k-\sigma)} \ .
\ee
Requiring that the brane does not bend gives
\be
\psi= \frac{2}{\sigma}\dz (\chi -\varphi)
\label{gauge1}
\ee
\be
v_\mu = \frac{1}{\sigma}\pa_\mu (\chi -\varphi) + B_\mu
\label{gauge2}
\ee
\be
\partial^\mu \bar \gamma_{\mu \nu} = 0 \ .
\label{gauge3}
\ee
Gauging away $B_\mu$, the remaining bulk equations (for $z>0$) reduce to
\be
\mu_0 \left(\pa_z^2 - 2(2k-\sigma)\pa_z + e^{2kz} \Box_4\right)
\bar \gamma_{\mu \nu} = 0
\label{BgE}
\ee
\be
4 \mu_0 (\omega+4/3) \left(\pa_z^2 - 2(2k-\sigma)\pa_z + e^{2kz} \Box_4\right)
\chi = 0
\label{BpE}
\ee
where $\mu_0=1-4\alpha(2\sigma-k)(\sigma-k)$. Again there is the
possibility of ghosts if the bulk curvature is too high. To avoid this
we need  $\mu_0 > 0$ (and $\omega > -4/3$).

Taking $\Lmat$ to be independent of $\Phi$, the brane junction
conditions imply
\be
2\mu_0\dz \bar \gamma_{\mu\nu}
+ \left(\beta - 8\alpha[2\sigma - k]\right) \Box_4 \bar \gamma_{\mu\nu}
=-\frac{2\sqrt{\alpha}}{\ms^2}\left\{S_{\mu\nu} - \frac{1}{3} 
\left(\eta_{\mu\nu} - \frac{\pa_\mu \pa_\nu}{\Box_4}\right) S \right\}
\label{bcgE}
\ee
\be
2\mu_0\dz \chi + \left(\beta - 8\alpha[2\sigma - k]\right) \Box_4 \chi
= -\frac{\sigma \sqrt{\alpha} }{(2\sigma-3k) \ms^2} S
\label{bcpE}
\ee
and
\be
\left(1+[4\alpha(2\sigma-k)-\beta][\sigma-k]\right) \Box_4 (\chi-\varphi)
=\frac{\sigma k \sqrt{\alpha}}{2(2\sigma-3k) \ms^2} S
\label{bcrE}
\ee
where $\ms^2 = M^3 e^{-2\phi_0} \sqrt{\alpha}$.

For spacelike momenta the solution of the bulk
equations~(\ref{BgE},\ref{BpE}) which vanishes as $z\to \infty$ is
\be
e^{(2k-\sigma)z} I_{2-\sigma/k} \left(-p e^{k z}/k\right) \ ,
\ee
unless $k = 0$ ($\omega=-1$), in which case it is
\be
\exp\left(-\left[\sigma +\sqrt{\sigma^2+p^2}\right]z\right) \ .
\ee

Substituting in the bulk solution and setting $z=0$, we find that the
left hand side of the graviton equation~\bref{bcgE} is equal to
\be
-\left\{ 2\mu_0 \frac{p I_{1-\sigma/k}(-p/k)}{I_{2-\sigma/k}(-p/k)}
+ (\beta - 16 \alpha [2+\omega] \sigma) p^2 \right\} \bar \gamma_{\mu\nu}
\label{gsolE}
\ee
for $\omega > -1$ and 
\be
-\left\{ 2\mu_0 (\sqrt{\sigma^2+p^2}+\sigma) 
+ (\beta - 16\alpha \sigma) p^2 \right\} \bar \gamma_{\mu\nu}
\label{gsolE1}
\ee
for $\omega=-1$. The left hand side of the scalar equation~\bref{bcpE}
will have a similar form.

If the above term in brackets vanishes for some $p>0$, then
eqs. \bref{bcgE} and \bref{bcpE} will have non-trivial solutions when
$\Lmat=0$. In this case the vacuum will have tachyon modes, and so
will be classically unstable. If the induced terms in the
action~\bref{actind} are absent, or small ($\beta < 16 \alpha
[2+\omega]$), this is always the case. Hence the addition of the
higher order gravity terms has destabilised the solution.

If the quadratic curvature terms are absent, the above brane world has
a negative tension brane, suggesting that the system is unstable. In
fact it is stable, but only because of the $Z_2$-symmetry. It is
therefore not at all surprising that adding extra terms to the action
(even if $\alpha$ is tiny) removes the stability. Similar effects were
noted for the brane world models discussed in ref.~\cite{us2}, and
also for the two brane Randall-Sundrum model~\cite{Dufaux}. The
problem could be avoided by taking $\beta$ to be large. This will be
discussed in section~\ref{sec:Ind}.

\section{Gauss-Bonnet Branches}
\label{sec:GB}

We will now consider the new branches of solutions which appear when
the quadratic curvature terms are included in the action. These have
the bulk, background solution
\be
k = -\omega \sigma \pm
 \sqrt{\frac{1}{12\alpha} - \frac{\omega(\omega+2)}{3}\sigma^2} \ .
\label{ksol}
\ee
We will only consider $-1 \leq \omega < 0$, since the field equations are
degenerate when $\omega=0$. As in the previous section
$\gamma_{\mu\nu}$ should be split into scalar and tensor parts. We use
the same gauge as before~(\ref{gauge1}--\ref{gauge3}), but this time take
\be
\gamma_{\mu \nu} = 
\bar \gamma_{\mu \nu} + \frac{1}{4} \gamma \eta_{\mu \nu} 
+\frac{4}{3}c_\chi \left(\frac{1}{4}\eta_{\mu \nu}
  -\frac{\pa_\mu \pa_\nu}{\Box_4}\right)\chi
\ee
where the field $\chi$ is now defined by
\be
\chi=\frac{3(8k\varphi - \sigma \gamma)}{4(6k+c_\chi\sigma )} \ ,
\ee
with
\be
c_\chi = 
\frac{\omega \sigma (3k+2(2\omega+1)\sigma)}{k(k+\omega \sigma)} \ .
\ee
Using the above expressions, the bulk equations reduce to
\be
\mu_\gamma \left(\dz^2 - 2(2k -\sigma)\dz 
+ f_\gamma^2 e^{2kz} \Box_4\right)\bar \gamma_{\mu \nu} = 0
\label{BgGB}
\ee
and
\be
\mu_\chi \left(\pa_z^2 - 2(2k -\sigma)\pa_z + 
f_\chi^2 e^{2kz} \Box_4\right) \chi = 0
\label{BpGB}
\ee
where
\be
f^2_\gamma = \frac{k-(1-\omega) \sigma}{\omega \sigma+k}
\ee
\be
f^2_\chi = \frac{3k}{3k+2(2\omega+1)\sigma}
\ee
\be
\mu_\gamma = 8\alpha(k+2[\omega+1]\sigma)(k+\omega\sigma)
\ee
\be
\mu_\chi = -16\omega\alpha (k+2[\omega+1]\sigma)(3k+2[2\omega+1]\sigma) \ .
\ee
As in sections \ref{sec:lin} and \ref{sec:Ein}, if the parameters
$\mu_\gamma$ and $\mu_\chi$ are not positive the solution will have
bulk ghosts.

In contrast to the Einstein gravity solutions of
section~\ref{sec:Ein} the tensor and scalar perturbations 
($\bar \gamma_{\mu \nu}$ and $\chi$ respectively) have different
behaviour in the bulk. For example, if $f_\gamma \ll f_\chi$, then the
tensor perturbations will be less localised on the brane than the
scalar perturbations. In further contrast to the bulk field equations of
the previous section, the parameters $f^2_\gamma$ and $f^2_\chi$ may be
negative. If this happens, the corresponding field equation will
resemble that for a field in a spacetime with three timelike
coordinates (or two if $\mu_{\gamma,\chi}$ is also negative). This
will give rise to instabilities on the quantum level and, for some parameter
ranges, on the classical level too. Either way, we need to have not
just $\mu_\gamma$ and $\mu_\chi$ positive, but also $f^2_\gamma$ and
$f^2_\chi$.

The brane junction conditions imply
\be
2\mu_\gamma \dz \bar \gamma_{\mu\nu}
+(\beta+ 8\alpha[k-2\sigma])  \Box_4 \bar \gamma_{\mu\nu}
=-\frac{2\sqrt{\alpha}}{\ms^2}\left\{S_{\mu\nu} 
- \frac{1}{3}\left(\eta_{\mu\nu} 
- \frac{\pa_\mu \pa_\nu}{\Box_4}\right)S \right\}
\label{bcgGB}
\ee
\be
\mu_\chi \left(2(k-\sigma)\dz \chi - f_\chi^2 \Box_4 \chi\right)
+2 k \left( [3+2\omega]\beta + 8\alpha[\omega+2][3k+2\omega\sigma]\right)
\Box_4 \varphi = -k\frac{\sqrt{\alpha}}{\ms^2}S
\label{bcpGB}
\ee
\be
\mu_\chi \sigma \dz \chi 
- 3 k (\beta + 16\alpha[k +\omega\sigma]) \Box_4 \chi 
+ (4+\beta[3k+2\omega \sigma])\sigma \Box_4 \varphi  = 0
\label{bcrGB}
\ee 
with $\ms^2 = M^3 e^{-2\phi_0} \sqrt{\alpha}$.
If the coefficient of the $\Box_4 \bar \gamma_{\mu\nu}$ term in the
boundary condition~\bref{bcgGB} is negative, the system may be
unstable. If $k<0$ the situation will be similar to that in the
previous section, and there will be a graviton tachyon mode. The same
will be true in $k>0$ and $\mu_\gamma$ is not too small. Alternatively
if $\mu_\gamma$ is sufficiently small there will be no tachyon, but the
effective four dimensional gravitational coupling will have the wrong
sign. Similar considerations apply to the scalar modes.

If we take the negative choice of sign for the solution~\bref{ksol},
then the requirement that there are no ghosts restricts $\sigma$ to be
less than  $\sigma_*=[8\alpha(2\omega^2+7\omega +6)]^{-1/2}$. In this
range $8\alpha[k-2\sigma] <0$ and $k$ is always negative. This branch
is therefore unstable (for small $\beta$). As with the solution in
section~\ref{sec:Ein} the instability could be removed by including a
large induced gravity term (i.e.\ taking $\beta>-8\alpha[k-2\sigma]$). 
We will consider this in section~\ref{sec:Ind}.

For the rest of this section we will take the positive sign in
eq.~\bref{ksol}.  For this branch, $k$, $\mu_\gamma$, $\mu_\chi$ and
$f_\chi^2$ are all positive, and $f_\chi^2 \geq f_\gamma^2$. For
$f_\gamma^2$ to be positive we require that $\sigma< \sigma_c$, where
$\sigma_c = [4\alpha(\omega^2+2\omega+3)]^{-1/2}$. To avoid tachyons
(graviton and scalar) we also need $\beta + 8\alpha(k-2\sigma) > 0$. 
If $\beta=0$ this is true when $\sigma \leq \sigma_*$. Since 
$\sigma_* \leq \sigma_c$, the junction conditions provide a stronger
constraint than the bulk effects.

The bulk graviton equation~\bref{BgGB} is solved by
\be
\bar\gamma_{\mu\nu}(p,z) \propto e^{(2k-\sigma)z} 
K_{2-\sigma/k} \left(f_\gamma p e^{k z}/k\right) \ .
\label{gsolGB}
\ee
The bulk scalar equation has a similar solution, but with $f_\gamma$
replaced by $f_\chi$.  The bulk graviton perturbations are
qualitatively similar those in the conventional brane world (see
section~\ref{sec:lin}). In fact we can get something similar to the
Randall-Sundrum model (with a scalar field and induced gravity term)
if we take $\sigma \to 0$. In particular $f_\chi$ and $f_\gamma$ will be
both equal to one.

The solution of the bulk field equation~\bref{gsolGB} implies that 
\be
\dz \bar\gamma_{\mu\nu} = - (p f_\gamma) 
\frac{K_{1-\sigma/k}(f_\gamma p/k)}{K_{2-\sigma/k}(f_\gamma p/k)} \, 
\bar\gamma_{\mu\nu}
\ee
on the brane. A similar relation holds for $\dz \chi$.  By using
asymptotic and series expansions of the above expression, we can
determine the approximate small and large distance effective
gravitational laws on the brane (just as we did in
section~\ref{sec:lin}). If $k > \sigma$ (as is the case for the
solutions we are considering), then when $p$ is small 
$\dz \bar\gamma_{\mu\nu} \approx  -f_\gamma^2 p^2 \bar\gamma_{\mu\nu}
/[2(k-\sigma)]$, and when $p$ is large 
$\dz \bar\gamma_{\mu\nu} \approx -f_\gamma p \, \bar\gamma_{\mu\nu}$.

At large distances, when $p \ll k/f_\chi$, the junction
conditions~(\ref{bcgGB}-\ref{bcrGB}) imply that the four dimensional
effective linearised gravity will be a tensor-scalar theory [see
eqs. \bref{BD1} and \bref{BD2}] with 
\be
\mpl^2 = \left(\beta + \frac{2-\mu_\gamma}{k-\sigma} \right)
\frac{\ms^2}{\sqrt{\alpha}}
\label{mpll}
\ee
\be
\mph^2 = \left[ (3+2\omega)(\beta/\sqrt{\alpha})
+ 8\sqrt{\alpha}(\omega+2)(3k+2\omega\sigma)\right] \ms^2 \ .
\label{mphl}
\ee
In this case the effective brane scalar field, $\tilde \varphi$, is
just $\varphi$.

To avoid conflict with experimental measurements, we would like to
have $\mpl \ll \mph$~\cite{Damour}. For $\omega=-1$ (so
$\sigma_*=\sigma_c$) and $\beta=0$ this will be true when
$f_\gamma \approx 0$, which occurs when $\sigma \approx \sigma_c$. We
then have $\mph \sim \ms$ and  $\mpl/\mph \sim f_\gamma$. Thus a
hierarchy between the effective scalar and tensor couplings has
arisen, despite the fact that the fundamental, five-dimensional scalar
and gravitational couplings are of the same order. This is
possible because the scalar field and the higher gravity terms have
affected the different parts of the perturbation in different
ways. This only occurs when the model includes both these features. If
$\Phi$ is turned off ($\sigma \to 0$), then $f_\gamma,f_\chi \to 1$
and both types of perturbation `feel' the effects of the bulk in the
same way. On the other hand if the higher gravity terms are dropped
($\alpha\to 0$), the above solution does not exist in the first place.

A similar effect can be obtained for other values of $\omega$, but
unfortunately the region of parameter space with $f_\gamma \approx 0$
($\sigma \approx \sigma_c$) has $8\alpha[k-2\sigma] <0$, which implies
that tachyons will be present.. This can be fixed by including an
appropriate induced gravity term with coefficient
$\beta = \beta_c = 8\alpha(1+\omega) \sigma_c$. A hierarchy is then
obtained, but at the price of a fine-tuning $\beta$.

At very short distances, when $p \gg k/f_\gamma$, the $\Box_4$ terms in
the boundary conditions dominate the $\dz$ terms. The resulting effective
gravitational theory is again scalar-tensor. The Planck mass is then
\be
\mpl^2 = [(\beta/\sqrt{\alpha})+ 8\sqrt{\alpha}(k-2\sigma)] \ms^2 \ .
\label{mpls}
\ee
This time the effective scalar mode is a combination of $\gamma$ and
$\varphi$. This gives a different relation between the two mass scales,
\be
\mph^2 = 3\mpl^2
\left[1+\omega\frac{\mpl^2}{\ms^2}\left(\frac{3\mpl^2}{2\ms^2} +
\frac{\omega \mu_\gamma}{k\sqrt{\alpha}} 
+ \frac{96\alpha(k+2[1+\omega]\sigma)}
{(\mpl/\ms)^2+16\sqrt{\alpha}(k+2[1+\omega]\sigma)}\right)^{\! -1}\right] \ .
\label{mphs}
\ee
For the parameter ranges we are considering, this gives
$\mph^2 \leq 3\mpl^2$, even when $\sigma$ is fine tuned to be near
$\sigma_c$. The suppression of the scalar part of the gravity is lost
at short distances. However the short distance constraints on
Brans-Dicke gravity are weak, so this is not a problem if the length
scale $f_\chi/k$ is of geographical size or smaller.

For $k/f_\gamma \gg p \gg k/f_\chi$, tensor perturbations have large
distance behaviour, while scalar perturbations have small distance
behaviour. This third regime is another new feature of this brane
world mode which is not present in the standard brane world. $\mpl$ is
again given by eq.~\bref{mpll} as above, but this time there is a different
effective scalar, and
\be
\mph^2 = 3\mpl^2\left[1 + 
\frac{\mpl^2 k(\sigma-k)}
{4\ms^2\sqrt{\alpha}(k+2[1+\omega]\sigma)\omega\sigma^2}
\left(2+\frac{3kY^2}{\sigma-k+\omega\sigma Y}\right)^{\! -1}\right]
\label{mphm}
\ee
where
\be
Y = 1+
\frac{\mpl^2(\sigma-k)}{8\ms^2\sqrt{\alpha}(k+2[1+\omega]\sigma)\omega\sigma}
\ .
\ee
As with the short distance gravity, we find $\mph^2 \leq 3\mpl^2$. As
before, this need not be a problem unless the length scale
$f_\gamma/k$ is of astronomical size.

The above expressions for $\mpl$ and $\mph$ at the three different scales
are algebraically rather complicated, and not particularly
revealing. It is instructive to take $\beta=\beta_c$,
$\sigma=\sigma_c-\sqrt{\alpha} \epsilon$, and then consider $\epsilon$
close to zero. In this case
\be
\mpl^2 \approx \frac{8(\omega+3)}{(-\omega)} \ms^2 \epsilon 
\label{Mplel}
\ee
at large and medium distances, and
\be
\mpl^2 \approx \frac{8(\omega+3)(\omega+2)}{3} \ms^2 \epsilon
\label{Mples}
\ee
at short distances. At medium and short distances, it can be seen from
eqs. \bref{mphs} and \bref{mphm} that $\mph^2$ is approximately 3
times the corresponding $\mpl^2$ (recall $\mpl \ll \ms$ when
$\beta=\beta_c$ and $\sigma\approx \sigma_c$). At large distances
\be
\mph^2
\approx \frac{4(\omega+3)^2}{\sqrt{\omega^2+2\omega+3}} \ms^2 \ .
\ee
We see that Brans-Dicke gravity constraints could be satisfied (at
large distance scales) by taking $\epsilon \lesssim 10^{-3}$. On the
other hand, we see from eqs. \bref{Mplel} and \bref{Mples} that there
is at least a factor of 3 difference between $\mpl^2$ on medium and
small distance scales. This may lead to conflict with observations.
Of course, any problems could be avoided by taking the
distance scale $f_\gamma/k$ to be very small. In this case gravity
on all scales at which accurate gravitational experiments have been
performed would be given by eqs. \bref{mpll} and \bref{mphl}.

\section{Scalar Gravity}
\label{sec:Ind}

We saw in section~\ref{sec:Ein} that if second order bulk curvature
terms were added to a brane world solution with a negative warp
factor, it could develop a tachyon mode. If a sufficiently large
induced gravity term is added to the action, this instability will be
removed.  Having included such a term, the approximate gravitational law
on the brane can be obtained (as with the other brane world solutions)
by substituting the solutions to the linearised bulk
equations~(\ref{gsolE},\ref{gsolE1}) into the junction
conditions~(\ref{bcgE},\ref{bcpE}).

At relatively small distances ($p \gg |k|$), we obtain a scalar-tensor theory
of gravity~(\ref{BD1},\ref{BD2}) with
\be
\mpl^2 = \left[(\beta/\sqrt{\alpha}) - 16(\omega+2)\sqrt{\alpha} \sigma\right]
\ms^2
\ee
and
\be
\mph^2 = (4+3\omega)\frac{\mu_0 - \sigma(3+2\omega)\mpl^2}
{(2+\omega)\mu_0 - \sigma(4+3\omega)\mpl^2} \mpl^2 \ .
\label{Mpind}
\ee
If $\omega=-1$ then $\mph =\mpl$. For other values of $\omega$ it
appears to be possible to obtain $\mph \gg \mpl$ by fine-tuning the
denominator of eq.~\bref{Mpind}, but unfortunately $\mph^2$ is negative in this
region of parameter space. Thus, for any choice of $\beta$, the above
theory will not be suitable for describing gravity in our universe at
astronomical distances.

Looking at the behaviour of the perturbations for small momenta (large
distances), we see that the junction conditions~(\ref{bcgE},\ref{bcpE}) 
imply that $\bar \gamma_{\mu\nu}, \chi \sim S$, while (for 
$\omega \neq -1$) $\varphi,\gamma \sim S/p^2$. To leading order in $p$
the tensor modes are negligible and so at large distances ($p \ll |k|$), 
we have a scalar theory of gravity.  A model with similar
behaviour was discussed in ref.~\cite{lingrav1}.

By considering the motion of a test particle which is confined to the
brane, we see that $d^2 x^i/dt^2 \approx -\hat \Gamma^i_{00}$, and so (for
static solutions) the gravitational potential is $V = (1/2)\gamma_{00}$. 
For the above scalar theory of gravity there is an order $1/p^2$
contribution to $V$ from $\gamma$, and so a Newtonian potential is
obtained. If on the other hand the particle is not confined to the
brane, $d^2 x^i/dt^2 \approx - \Gamma^i_{00} = -\hat \Gamma^i_{00}+ k v_i$. 
Using the expression for $v_i$~\bref{gauge2}, we find that now
only $\bar \gamma_{00}$ and $\chi$ contribute to $V$, and so no
Newtonian potential is obtained. Therefore any Newtonian potential (at
large distances) is purely a result of the brane's curvature, and
nothing to do with the bulk behaviour of the perturbations.

The large distance behaviour of the above model is also incompatible with
astronomical observations. The addition of an induced gravity term to
the solutions of section~\ref{sec:Ein} may have solved the stability
problem, but it still does not produce a theory which could
represent our universe. If $\omega =-1$, we will instead have a
non-standard, scalar-tensor, large distance gravitational law, with a
non-Newtonian gravitational potential. We will leave the analysis of
such a theory for future work.

The `Gauss-Bonnet' solutions in section~\ref{sec:GB} with the negative
sign choice in eq.~\bref{ksol} suffered from a similar instability to
the above `Einstein' solution. This can also be cured by adding a
large induced gravity term. In general, these stabilised solutions
will also behave as scalar gravity theories with Newtonian
gravitational potentials at large distances. At short distances they
will be approximately Brans-Dicke, with $\mpl$ and $\mph$ given by
eqs. \bref{mpls} and \bref{mphs}. In contrast to the above `Einstein'
solution, it does appear to be possible to fine-tune $\beta$ so that
$\mph \gg \mpl$ without having $\mph<0$ (although we will not discuss
the algebraic details here). In this case a hierarchy is obtained by
fine-tuning the positive contribution to the scalar coupling from the
induced action~\bref{actind}, so that it cancels the negative
contribution from the other boundary terms~\bref{act2}. This contrasts
with the hierarchy obtained in section~\ref{sec:GB}, which arose from
the properties of the bulk background solution.

As with the other `Gauss-Bonnet' solutions discussed in
section~\ref{sec:GB}, these solutions also have a medium distance
gravitational law which is different from the small and large distance
laws. Since this solution branch has $f_\chi \leq f_\gamma$, the
tensor modes will have short distance behaviour in this region, and
the scalar modes will have large distance behaviour (the reverse of
what happens for the other `Gauss-Bonnet' solutions). Thus for
$k/f_\chi \gg p \gg k/f_\gamma$ we will again have scalar-tensor gravity, but
with different effective mass scales to the short distance theory.

\section{Conclusions}

In this paper we have investigated the effects of higher order
curvature terms and a scalar field on linearised gravity in a brane
world scenario. These two extensions of the conventional brane world give rise
to many new features.  The higher order gravity produces additional terms
in the brane junction conditions. These resemble those produced by an
induced gravity term, and allow gravity to be four dimensional at all
length scales, weakening constraints on the model. Another, less desirable,
consequence is the possibility of new instabilities. If the curvature
is too high, corrections to the bulk equations result in the
appearance of ghosts. If the extra contribution to the junction
conditions has the wrong sign, the model will have tachyons.

Not surprisingly, the gravitational field equations for the
quadratic theory have more solutions than the linear one. One of the
two extra solution branches gives a better behaved brane world than the
one obtained from the linear theory. In particular it is stable, and
its bulk space is warped in a similar way to the Randall-Sundrum
model. The other solution branches are unstable when the higher order
gravity effects are included.

The introduction of the scalar field alters how perturbations `feel'
the warping of the bulk spacetime. The various coefficients in the
linearised field equations are altered by presence of the bulk scalar
field and the higher gravity terms. Significantly the changes to
the scalar and tensor perturbation equations are not the same, and so
with some fine-tuning of the background solution, it is possible to
suppress the scalar modes relative to the tensor modes. This does not
happen unless we include both the scalar field {\em and} the higher
gravity terms. The effective four-dimensional theory on the
brane is Brans-Dicke gravity which, thanks to the suppression of the scalar
modes, could be compatible with observational constraints. However it
should be noted that the constraints on Brans-Dicke gravity were
obtained by going beyond a linearised approximation of the
theory. Since we have only considered a linearised theory in this
paper, the actual constraints may be different.

The instabilities in the other solution branches can (in some cases)
be removed by including sufficiently large induced gravity terms in
the brane action. The resulting theory is then a standard
four-dimensional gravity action, plus the brane world part of the
model. Since Brans-Dicke theory is stable, it is not too
surprising that the resulting combined theory will also be stable if
the brane world part of it is small. The resulting brane gravity is
approximately scalar-tensor at short distances. It appears that the
scalar modes can be suppressed in some cases by tuning the size of the
induced term to cancel the brane world contribution. A Newtonian
potential is obtained at large distances, although (to leading order)
the resulting theory has no tensor part.

Having included quadratic curvature terms in the theory it is very
natural to ask whether even higher order terms should be included as
well. If the solutions in this paper had small curvature, we could
safely ignore higher terms. Inevitably this is not the case, since if
the curvature were small, the effects of the quadratic terms would
also be small, and our brane world solutions would not be
significantly different from the usual brane world scenarios with
scalar fields. Work on brane worlds with higher order gravity
actions~\cite{Meissner} (without scalar fields) suggests that these
higher curvature terms will give qualitatively similar contributions
to the  linearised gravity as the quadratic ones. Hence the
results obtained in this paper should remain qualitatively unchanged,
although obviously the algebraic expressions for the different mass and length
scales in the models will be different.

\section*{Acknowledgements}

I wish to thank Phillippe Brax, Christos Charmousis, Justin Khoury,
Valery Rubakov and Slava Rychkov for useful comments and
discussions. I am grateful to the EU network HPRN--CT--2000--00152 and
the Swiss Science Foundation for financial support.

\appendix
\section{Appendix: Field Equations and Boundary Terms}

Variations of the various terms in the actions \bref{act1} and
\bref{act2} are given below. The results are derived for a
$D$-dimensional manifold, $\mathcal{M}$, bounded by a hypersurface
$\Sigma$. The non-standard convention that the normal of $\Sigma$
points into $\mathcal{M}$ is used, so 
$\int_\mathcal{M} d^Dx \sqrt{-g} \, \nabla^a F_a 
= - \int_\Sigma d^{D-1}x \sqrt{-h} \, n^a F_a$ (for a brane world,
this is slightly easier to visualise). Note that there is an extra
factor of two in the brane world boundary terms, since the brane is
treated as the boundaries of each of the two bulk
half-spacetimes. In the expressions below $\xi$ is an arbitrary
function. In the rest of the paper it was equal to $e^{-2\Phi}$. 
\bea
&&
\delta\left\{ \int_\mathcal{M} d^Dx \sqrt{-g} \, \xi R
-2\int_\Sigma d^{D-1}x \sqrt{-h} \, \xi K\right\}
\nonumber \\ && {}
= \int_\mathcal{M} d^Dx \sqrt{-g} \, \delta g^{ab} 
\left\{ \xi G_{ab}
- \nabla_{\! a} \nabla_{\! b}\xi + g_{ab}\nabla^2\xi \right\}
+\delta \Phi \{ \partial_\Phi \xi R \}
\nonumber \\ && {}
-\int_\Sigma d^{D-1}x \sqrt{-h} \, \delta g^{ab}\left\{ \xi(K_{ab}- K h_{ab})
- \partial_n\xi h_{ab}\right\} + \delta \Phi \{2 \partial_\Phi \xi K\}
\eea
\bea
&&
\delta \int_\mathcal{M} d^Dx \sqrt{-g} \, \xi (\nabla\Phi)^2 
\nonumber \\ && {}
= \int_\mathcal{M} \! d^Dx \sqrt{-g} \, \delta g^{ab}\left\{ \xi \nabla_{\! a}
\Phi \nabla_{\! b}\Phi - \frac{\xi}{2} g_{ab}(\nabla\Phi)^2 \right\}
+\delta \Phi\left\{\partial_\Phi \xi(\nabla\Phi)^2 
- 2\nabla^a \xi \nabla_{\! a} \Phi- 2 \xi \nabla^2 \Phi\right\}
\nonumber \\ && {}
-\int_\Sigma d^{D-1}x \sqrt{-h}\, \delta\Phi\left\{2 \xi \partial_n\Phi\right\}
\eea
Before dealing with the quadratic curvature terms, it is useful to
define the tensors
\be
J_{ab} = \frac{1}{3}\left(2K K_{ac}K^c{}_b + K_{cd}K^{cd} K_{ab} 
- 2K_{ac}K^{cd}K_{db} - K^2 K_{ab} \right) \ ,
\ee
\be 
P_{abcd} = R_{abcd} + 2 R_{b[c} g_{d]a} - 2 R_{a[c} g_{d]b} + R g_{a[c}g_{d]b} 
\ ,
\ee
so $\LGB = P^{abcd} R_{abcd}$. The second order Lovelock
Tensor~\cite{Lovelock} is then 
\be
H_{ab} = P_a{}^{cde} R_{bcde} - \frac{1}{4}g_{ab}\LGB \ .
\ee
\bea
&&
\delta\left\{ \int_\mathcal{M} d^Dx \sqrt{-g} \, \xi
\LGB - \int_\Sigma d^{D-1}x \sqrt{-h}
\, 4\xi (J - 2\widehat G^{ab}K_{ab}) \right\}
\nonumber \\ && {}
= \int_\mathcal{M} d^Dx \sqrt{-g} \, \delta g^{ab} \left\{ 2\xi H_{ab}
+ 4P_{acbe} \nabla^e \nabla^c \xi
\right\} + \delta \Phi \{ \partial_\Phi \xi \LGB \}
\nonumber \\ && {}
-\int_\Sigma d^{D-1}x \sqrt{-h} \, \delta g^{ab}\biggl\{\xi 
\left(6 J_{ab} -2J h_{ab} - 4\widehat P_{acbe} K^{ce} \right)
\nonumber \\ && \hspace{.6in} {}
+ 2 \partial_n \xi \left(2\widehat G_{ab} + 2 K_{ea}K^e{}_b - 2 K K_{ab}  + 
h_{ab} [K^2 - K_{cd}K^{cd}]\right) 
\nonumber \\ && \hspace{.6in} {}
+ 4\left(2K_{c(a}D^c D_{b)} \xi - K D_a D_b \xi -K_{ab}D^2\xi
- h_{ab}[K_{ec}D^eD^c\xi-K D^2\xi] \right) \biggr\}
\nonumber \\ && \hspace{.4in} {}
+ \delta \Phi \left\{ 4\partial_\Phi \xi (J - 2\widehat G^{ab}K_{ab})\right\}
\eea
\bea
&&
\delta\left\{ \int_\mathcal{M} d^Dx \sqrt{-g} \, \xi
G_{ab} \nabla^a\Phi \nabla^b\Phi - \int_\Sigma d^{D-1}x \sqrt{-h}
\, \xi (K_{ab} - K h_{ab}) D^a \Phi D^b \Phi\right\}
\nonumber \\ && {}
= \int_\mathcal{M} d^Dx \sqrt{-g} \, \delta g^{ab} \biggl\{ \xi\Bigl(
P_{acbe} \nabla^c\Phi \nabla^e\Phi + \frac{1}{2} G_{ab} (\nabla\Phi)^2
+\nabla_{\! c} \nabla_{\! (a} \Phi\nabla_{\! b)} \nabla^c \Phi 
-\nabla_{\! a} \nabla_{\! b}\Phi \nabla^2\Phi \Bigr)
\nonumber \\ && \hspace{.6in}{}
+\nabla^c \Phi \nabla_{\! (a} \xi\nabla_{\! b)} \nabla_{\! c} \Phi
+\nabla^c \xi \nabla_{\! (a} \Phi\nabla_{\! b)} \nabla_{\! c} \Phi
-\nabla^2\Phi \nabla_{\! (a} \Phi \nabla_{\! b)}\xi
-\nabla^c\Phi \nabla_{\! c} \xi \nabla_{\! a} \nabla_{\! b}\Phi
\nonumber \\ && \hspace{.6in}{}
- \nabla_{\! (a} \Phi \nabla_{\! b)} \nabla_{\! c} \xi  \nabla^c\Phi
+\frac{1}{2}\left(\nabla_{\! a} \Phi \nabla_{\! b}\Phi \nabla^2 \xi
+(\nabla\Phi)^2 \nabla_{\! a} \nabla_{\! b} \xi \right)
\nonumber \\ && \hspace{.6in}{}
+\frac{1}{2}g_{ab}\Bigl[
\xi(\nabla^2\Phi)^2 - (\nabla\Phi)^2 \nabla^2 \xi
+2\nabla^c\xi \nabla_{\! c} \Phi \nabla^2\Phi 
\nonumber \\ && \hspace{1.2in}{}
- \xi \nabla_{\! c} \nabla_{\! e} \Phi \nabla^c\nabla^e\Phi
+\nabla^c \Phi \nabla^e\Phi \nabla_{\! c} \nabla_{\! e} \xi 
- 2\nabla^c \xi \nabla^e \Phi\nabla_{\! c} \nabla_{\! e} \Phi
\Bigr] \biggr\}
\nonumber \\ && \hspace{.4in} {}
+ \delta \Phi G_{ab}\left\{\partial_\Phi\xi \nabla^a\Phi \nabla^b\Phi
-2\xi\nabla^a\nabla^b\Phi - 2\nabla^a\xi\nabla^b\Phi \right\} 
\nonumber \\ && {}
-\int_\Sigma d^{D-1}x \sqrt{-h} \, \delta g^{ab}\biggl\{\xi \Bigl( 
\frac{1}{2} [(\partial_n \Phi)^2-(D\Phi)^2] [K_{ab}- h_{ab}K]
+\partial_n \Phi (D_a D_b \Phi-h_{ab} D^2\Phi)
 \nonumber \\ &&  \hspace{.6in}{} 
+ 2K_{c(a} D^c \Phi D_{b)} \Phi - K D_a \Phi D_b \Phi
- h_{ab} K^{ec} D_e \Phi D_c \Phi\Bigr) 
 \nonumber \\  && \hspace{.6in} {}
+\partial_n \Phi \left[D_{(a} \Phi D_{b)} \xi  - h_{ab} D^c\Phi D_c\xi\right]
-\frac{1}{2}\partial_n \xi \left[D_a \Phi D_b \Phi  - h_{ab}(D\Phi)^2 \right]
\biggr\}
\nonumber \\  && \hspace{.4in} {}
+\delta \Phi \biggl\{ \xi (K^2-K^{ab}K_{ab}-\widehat R) \partial_n\Phi
\nonumber \\  && \hspace{.9in} {}
+ \left(\partial_\Phi \xi D^a \Phi D^b \Phi 
- 2 D^a \xi D^b \Phi - 2\xi D^a D^b \Phi\right) 
\left(K_{ab} - K h_{ab}\right) \biggr\}
\eea
\bea
&&
\delta\left\{ \int_\mathcal{M} d^Dx \sqrt{-g} \, \xi
(\nabla\Phi)^2 \nabla^2 \Phi 
+\int_\Sigma d^{D-1}x \sqrt{-h} \, \xi \partial_n \Phi \left(
\frac{1}{3}(\partial_n\Phi)^2 + (D\Phi)^2\right)\right\}
\nonumber \\ && {}
= \int_\mathcal{M} d^Dx \sqrt{-g} \, \delta g^{ab} \biggl\{ \xi  \left[
\nabla_{\! a} \Phi \nabla_{\! b}\Phi \nabla^2\Phi
-2\nabla_{\! c} \Phi \nabla_{\! (a} \Phi\nabla_{\! b)} \nabla^c \Phi
+g_{ab}\nabla_{\! c} \nabla_{\! e} \Phi \nabla^c\Phi \nabla^e\Phi\right]
\nonumber \\ && \hspace{1.4in} {}
+\frac{1}{2} (\nabla\Phi)^2\left[
g_{ab} \nabla_{\! c}\xi \nabla^c\Phi 
- 2 \nabla_{\! (a} \xi \nabla_{\! b)}\Phi\right] \biggr\}
\nonumber \\ && \hspace{.4in} {}
\delta \Phi \Bigl\{ 2\xi[R_{ab}\nabla^a\Phi \nabla^b\Phi+
\nabla_{\! a} \nabla_{\! b} \Phi \nabla^a\nabla^b\Phi - (\nabla^2\Phi)^2]
+4\nabla_{\! a} \xi \nabla_{\! b} \Phi \nabla^a \nabla^b\Phi
\nonumber \\ && \hspace{.6in} {}
+(\partial_\Phi \xi (\nabla\Phi)^2 
 -2\nabla_{\! a} \xi \nabla^a \Phi) \nabla^2\Phi
+\nabla^2 \xi (\nabla\Phi)^2 \Bigr\}
\nonumber \\ && {}
-\int_\Sigma d^{D-1}x \sqrt{-h} \, \delta g^{ab} \left\{-\xi\partial_n\Phi 
\left[\frac{1}{3} (\partial_n\Phi)^2 h_{ab} + D_a \Phi D_b \Phi\right]\right\}
\nonumber \\ && \hspace{.4in} {}
+\delta \Phi \biggl\{ 2\xi\left( 2 \partial_n\Phi D^2\Phi 
+ K (\partial_n\Phi)^2+  K^{ec} D_e \Phi D_c \Phi\right)
+2 \partial_n \Phi D^a \xi D_a \Phi 
\nonumber \\ && \hspace{.9in} {}
-\partial_\Phi \xi \partial_n \Phi 
\left( \frac{1}{3}(\partial_n\Phi)^2 + (D\Phi)^2\right)
- \partial_n \xi [(\partial_n\Phi)^2 + (D\Phi)^2]
 \biggr\}
\eea
\bea
&&
\delta\left\{ \int_\mathcal{M} d^Dx \sqrt{-g} \, \xi (\nabla\Phi)^4 \right\}
\nonumber \\ && {}
= \int_\mathcal{M} d^Dx \sqrt{-g} \, \delta g^{ab} \biggl\{ 
\xi  (\nabla \Phi)^2 \left( 2\nabla_{\! a} \Phi \nabla_{\! b}\Phi 
-\frac{1}{2}g_{ab}(\nabla \Phi)^2 \right) \biggr\}
\nonumber \\ && \hspace{.4in} {}
+\delta\Phi \left\{\partial_\Phi \xi (\nabla \Phi)^4
- 4\nabla_{\! a}\xi \nabla^a\Phi(\nabla \Phi)^2 
-4\xi (\nabla \Phi)^2 \nabla^2 \Phi 
-8 \xi \nabla_{\! a} \nabla_{\! b} \Phi \nabla^a\Phi\nabla^b\Phi
\right\}
\nonumber \\ && {}
-\int_\Sigma d^{D-1}x \sqrt{-h} \, \delta\Phi \left\{4\xi\partial_n\Phi
\left[(\partial_n\Phi)^2+(D\Phi)^2\right]\right\}
\eea

\end{document}